\documentclass[12pt]{article}
\usepackage{times}
\usepackage{fullpage}

\usepackage{graphicx}
\usepackage{amssymb}
\usepackage{amsmath}
\usepackage{ulem}
\usepackage[numbers]{natbib}
\usepackage[labelfont=bf]{caption}

\newcommand{\etacar}{$\eta$~Car}
\newcommand{\NUS}{{\it NuSTAR}}
\newcommand{\ARCSEC}{$''$}
\newcommand{\ARCMIN}{$'$}
\newcommand{\FOV}{{\it FOV}}
\newcommand{\PSF}{PSF}
\newcommand{\DEGREE}{{$^{\circ}$}}
\newcommand{\LX}{$L_{\rm X}$}
\newcommand{\UNITSOLARMASS}{$M_{\odot}$}
\newcommand{\UNITFLUX}{{\rm ergs~cm$^{-2}$~s$^{-1}$}}
\newcommand{\UNITSOLARABUND}{{\it Z$_{\odot}$}}
\newcommand{\UNITLUMI}{{\rm ergs~s$^{-1}$}}
\newcommand{\UNITVEL}{{\rm km~s$^{-1}$}}
\newcommand{\Mdot}{{$\dot M$}}
\newcommand{\INTEGRAL}{{\it INTEGRAL}}
\newcommand{\SUZAKU}{{\it Suzaku}}
\newcommand{\FERMI}{{\it Fermi}}
\newcommand{\AGILE}{{\it AGILE}}
\newcommand{\KT}{{\it kT}}
\newcommand{\NH}{{\it N$_{\rm H}$}}
\newcommand{\UNITNH}{{\rm cm$^{-2}$}}
\newcommand{\UNITSOLARLUMI}{{\it L$_{\odot}$}}
\newcommand{\CHANDRA}{{\it Chandra}}
\newcommand{\XMM}{{\it XMM-Newton}}
\newcommand{\SWIFT}{{\it Swift}}
\newcommand{\RXTE}{{\it RXTE}}
\newcommand{\apj}{{\it Astrophys. J.}}
\newcommand{\apjl}{{\it Astrophys. J. Lett.}}
\newcommand{\apjs}{{\it Astrophys. J. Suppl.}}
\newcommand{\aj}{{\it Astron. J.}}
\newcommand{\aap}{{\it Astron. Astrophys.}}
\newcommand{\mnras}{{\it Mon. Not. R. Astron. Soc.}}

\newcommand{\pasj}{{\it Publ. Astron. Soc. Jpn}}
\newcommand{\nat}{{\it Nature}}

\def\@cite#1#2{$^{\mbox{\scriptsize #1\if@tempswa , #2\fi}}$}

\newcommand{\spacing}[1]{\renewcommand{\baselinestretch}{#1}\large\normalsize}
\spacing{2}

\def\@maketitle{%
  \newpage\spacing{1}\setlength{\parskip}{12pt}%
    {\Large\bfseries\noindent\sloppy \textsf{\@title} \par}%
    {\noindent\sloppy \@author}%
}

\newenvironment{affiliations}{%
    \setcounter{enumi}{1}%
    \setlength{\parindent}{0in}%
    \slshape\sloppy%
    \begin{list}{\upshape$^{\arabic{enumi}}$}{%
        \usecounter{enumi}%
        \setlength{\leftmargin}{0in}%
        \setlength{\topsep}{0in}%
        \setlength{\labelsep}{0in}%
        \setlength{\labelwidth}{0in}%
        \setlength{\listparindent}{0in}%
        \setlength{\itemsep}{0ex}%
        \setlength{\parsep}{0in}%
        }
    }{\end{list}\par\vspace{12pt}}

\renewenvironment{abstract}{%
    \setlength{\parindent}{0in}%
    \setlength{\parskip}{0in}%
    \bfseries%
    }{\par\vspace{-6pt}}

\newenvironment{addendum}{%
    \setlength{\parindent}{0in}%
    \small%
    \begin{list}{Acknowledgements}{%
        \setlength{\leftmargin}{0in}%
        \setlength{\listparindent}{0in}%
        \setlength{\labelsep}{0em}%
        \setlength{\labelwidth}{0in}%
        \setlength{\itemsep}{12pt}%
        }
    }
    {\end{list}\normalsize}

\title{Non-thermal X-rays from Colliding Wind Shock Acceleration in the Massive Binary $\eta$ Carinae}

\author{Kenji Hamaguchi$^{1,2, *}$, 
Michael F. Corcoran$^{1,3}$,
Julian M. Pittard$^{4}$,\\
Neetika Sharma$^{2}$,
Hiromitsu Takahashi$^{5}$,
Christopher M. P. Russell$^{6,7}$,\\
Brian W. Grefenstette$^{8}$,
Daniel R. Wik$^{9}$,
Theodore R. Gull$^{6}$,\\
Noel D. Richardson$^{10}$,
Thomas I. Madura$^{11}$, \&
Anthony F. J. Moffat$^{12}$
}

\date{\it Nature Astronomy 2 (2018) 731-736}

\begin{document}

\maketitle

\begin{affiliations}
	\item CRESST II and X-ray Astrophysics Laboratory NASA/GSFC, Greenbelt, MD 20771, USA, $^{*}$Kenji.Hamaguchi@nasa.gov
	\item Department of Physics, University of Maryland, Baltimore County, 1000 Hilltop Circle, Baltimore, MD 21250, USA
	\item The Catholic University of America, 620 Michigan Ave. N.E., Washington, DC 20064, USA
	\item School of Physics and Astronomy, The University of Leeds, Woodhouse Lane, Leeds LS2 9JT, UK
	\item Department of Physical Sciences, Hiroshima University, Higashi-Hiroshima, Hiroshima 739-8526, Japan
	\item Astrophysics Science Division, NASA Goddard Space Flight Center, Greenbelt, MD 20771, USA
	\item Instituto de Astrof\'{i}sica, Pontificia Universidad Cat\'{o}lica de Chile, Santiago, Chile
	\item Space Radiation Lab, California Institute of Technology, Pasadena, CA 91125, USA
	\item Department of Physics \& Astronomy, University of Utah, Salt Lake City, UT 84112, USA
	\item Ritter Observatory, Department of Physics and Astronomy, The University of Toledo, Toledo, OH 43606-3390, USA
	\item San Jose State University, Department of Physics \& Astronomy, One Washington Square, San Jose, CA, 95192-0106, USA
	\item D\'epartement de physique and Centre de Recherche en Astrophysique du Qu\'ebec (CRAQ), Universit\'e de Montr\'eal, C.P. 6128, Canada
\end{affiliations}

\begin{abstract}
Cosmic-ray acceleration has been a long-standing mystery \cite{Koyama1995a,Morlino2012a} 
and despite more than a century of study, we still do not have a complete census of acceleration mechanisms.
The collision of strong stellar winds in massive binary systems creates powerful shocks,
which have been expected to produce high-energy cosmic-rays through Fermi acceleration at the shock interface.
The accelerated particles should collide with stellar photons or ambient material,
producing non-thermal emission observable in X-rays and $\gamma$-rays \citep{Pittard2006,DeBecker2017}.
The supermassive binary star \etacar\ drives the strongest colliding wind shock in the solar neighborhood \citep{Corcoran2005,Groh2012}.
Observations with non-focusing high-energy observatories indicate 
a high energy source near \etacar, but have been unable to conclusively identify \etacar\ as the source
because of their relatively poor angular resolution \citep{Leyder2008,Sekiguchi2009,Abdo2010}.
Here we present the first 
direct focussing observations of the non-thermal source in the extremely hard X-ray band,
which is found to be spatially coincident with the star within several arc-seconds.
These observations show that the source of non-thermal X-rays varies with the orbital phase of the binary,
and that the photon index of the emission is similar to that derived through analysis of the $\gamma$-ray spectrum.
This is conclusive evidence that the high-energy emission indeed originates from non-thermal particles
accelerated at colliding wind shocks.
\end{abstract}

Strong shocks accelerate particles to cosmic-ray energies through Fermi acceleration.
Supernova remnants are well established as a source of cosmic rays in the Milky Way \citep{Koyama1995a,Morlino2012a},
but other sources may also contribute.
Massive, luminous hot stars drive powerful stellar winds through their UV radiation \citep{Castor1975} 
and, in a massive binary system,
the collision of the stellar winds will produce strong shocks and thermal X-ray emission. 
This wind-wind collision region may serve as an additional source of cosmic-ray particles. 
Indeed, non-thermal radio emission from colliding wind binary systems is often detected
\citep{Dougherty2000a,DeBecker2013}, and has been directly imaged by high-spatial-resolution observations 
\citep[e.g.,][]{Williams1997,Dougherty2005}.
The emission is interpreted as radio synchrotron emission from high energy non-thermal electrons.
These accelerated, non-thermal particles can also produce high energy X-ray and $\gamma$-ray photons 
through inverse-Compton (IC) scattering of stellar UV photons or pion-decay after collision with ambient material.
However, the detection of high energy non-thermal X-ray and $\gamma$-ray emission from colliding wind binaries is currently very challenging, and the handful of reported detections remain controversial \citep[see, e.g.,][]{DeBecker2017}.

The best candidate massive binary system for detecting the high-energy non-thermal radiation produced by 
a shock-accelerated population of high-energy particles is \etacar.
Eta Carinae is the most luminous binary in our Galaxy and the variable thermal X-ray emission
produced by the hot plasma (\KT~$\sim$4$-$5~keV, \LX $\sim$10$^{35}$~\UNITLUMI) in its colliding wind shock has been well studied 
\citep[][and references therein]{Corcoran2017a}.
The primary is one of the most massive stars in our Galaxy \citep[$\gtrsim$100~\UNITSOLARMASS,][]{Hillier2001}
and drives a powerful wind \citep[$v\sim$ 420~\UNITVEL, \Mdot~$\sim$8.5$\times$10$^{-4}$\UNITSOLARMASS~yr$^{-1}$,][]{Groh2012}.
The secondary is perhaps a massive star of O or Wolf-Rayet type, which has never been directly observed, though its wind properties 
\citep[$\sim$3000~\UNITVEL, \Mdot~$\sim$10$^{-5}$\UNITSOLARMASS~yr$^{-1}$,][]{Pittard2002} 
have been deduced through analysis of its X-ray spectrum.
Variations across the electromagnetic spectrum from \etacar\ have shown that the system has a long-period orbit with high eccentricity
\citep[$e \sim$0.9, $P \sim$5.54 yrs,][]{Corcoran2005,Damineli2008}.

In extremely high energy X-rays (15$-$100~keV),
the \INTEGRAL\ and \SUZAKU\ observatories claimed detection of a non-thermal source near \etacar\
\citep{Leyder2008,Leyder2010,Sekiguchi2009,Hamaguchi2014b}, 
but two more sensitive \NUS\ observations near periastron in 2014 did not confirm this \citep{Hamaguchi2016a}.
The \AGILE\ and \FERMI\ space observatories detected a GeV $\gamma$-ray source near \etacar\ \citep{Tavani2009,Abdo2010},
while the 
HESS telescope detected a source of high-energy $\gamma$-ray emission \citep{Leser2017a} at energies up to 300 GeV.
The $\gamma$-ray spectrum shows two components, above and below 10~GeV.
Both components vary slowly with {\etacar}'s orbital phase \citep[e.g.,][]{Reitberger2015}.
The poor angular resolutions ($\gtrsim$10$'$) of these observations
meant that \etacar\ could not be conclusively confirmed as the source of the high-energy emission.

\begin{figure}
\includegraphics[width=15cm]{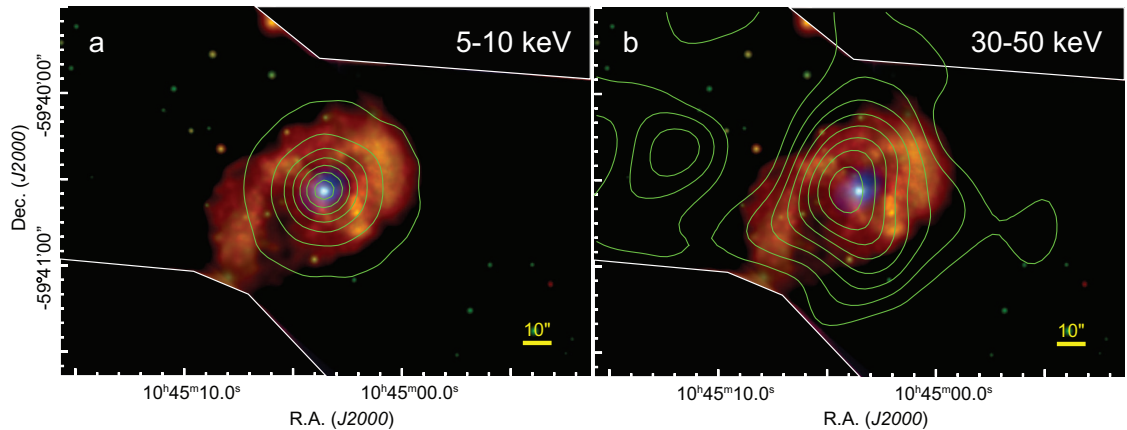}
\caption{
\textbf{\NUS\ image contours of the \etacar\ field.}
The contours in a conventional X-ray band (5$-$10~keV, {\bf a}) and an extremely hard X-ray band (30$-$50 keV, {\bf b})
are produced from the \NUS\ observations on 2015 July 16 ($\phi_{\rm orb} =$0.17) and 2016 June 15  ($\phi_{\rm orb} =$0.34)
and overlaid on a true colour X-ray image of the same field taken with the \CHANDRA\ X-ray observatory 
during the soft X-ray minimum in 2009 \citep{Hamaguchi2014a}.
The contours are drawn at intervals of 10\% starting from the X-ray peak above background.
The \NUS\ images were aligned with the \CHANDRA\ image by matching the peak of the thermal emission at $E <$10~keV 
in the \NUS\ image with that of the \CHANDRA\ image.
The 30$-$50 keV source centroid, which has an uncertainty of about 5\ARCSEC\ at 2$\sigma$,
is consistent with the centroid of the thermal, 5$-$10 keV source (i.e., \etacar).
Earlier measurements of extremely hard X-ray and $\gamma$-ray source positions are 
constrained at an accuracy of $\sim$1\ARCMIN\ or larger \citep[e.g.,][]{Leyder2010,Reitberger2015}.
\label{fig:image}
}
\end{figure}

The \NUS\ X-ray observatory, launched in 2012,
provides for the first time focusing observations at energies up to 79~keV \citep{Harrison2013}.
We obtained 11 \NUS\ observations of \etacar\ around \etacar's last periastron passage in 2014
through 2015 and 2016, along with coordinated observations at energies between 0.3$-$12~keV with the \XMM\ observatory \cite{Jansen2001}.
The \NUS\ image at the highest available energy in which the source can be detected above background (30$-$50~keV) shows, 
for the first time, that even at these high energies the emission clearly arises 
in the direction of and is well-centered on the position of \etacar\ (Figure~\ref{fig:image}).

\begin{figure}
\includegraphics[width=15cm]{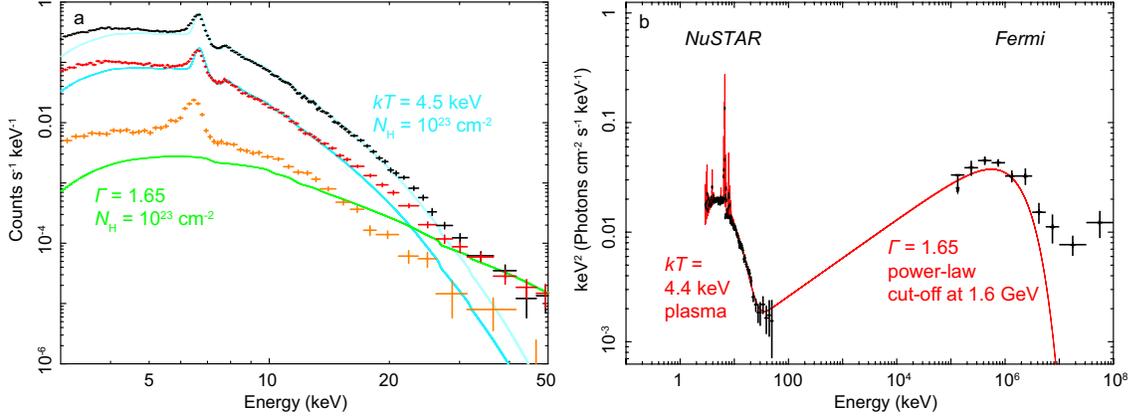}
\caption{
\textbf{\NUS\ spectra in three characteristic orbital phases of \etacar\ and a comparison 
to a \FERMI\ $\gamma$-ray spectrum.}
{\bf a}, \NUS\ spectra obtained during the rise of the soft X-ray flux toward periastron on 2014 March 31 ({\it black}, $\phi_{\rm orb} =$0.94),
the soft X-ray minimum on 2014 August 11 ({\it orange}, $\phi_{\rm orb} =$0.005), and
after the soft X-ray flux recovery from the 2014 periastron event ({\it red}).
The last spectrum is co-added from two spectra
in 2015 July 16 ($\phi_{\rm orb} =$0.17) and 2016 June 15 ($\phi_{\rm orb} =$0.34), to increase the signal-to-noise.
The vertical axis shows the raw photon counts from the detector.
Error bars are shown at 1$\sigma$.
The cyan and green solid lines show emission of \KT\ = 4.5 keV thermal plasma and a $\Gamma$ = 1.65 power-law,
which are convolved with the detector response, to give expected histograms of the detector counts at each energy.
The thin cyan spectrum is $\sim$4 times brighter than the thick cyan spectrum.
The excess from the \KT\ = 4.5 keV thermal plasma emission below $\sim$6~keV mostly 
originates from a lower temperature (\KT\ $\sim$1.1~keV) component.
{\bf b}, \NUS\ spectrum on 2016 June 15 and a \FERMI\ spectrum \citep{Abdo2010}
after correcting the detector response ({\it black})
compared to the best-fit spectral model, a $\Gamma$ = 1.65 power-law cut-off at 1.6 GeV ({\it red}).
\label{fig:spectra}
}
\end{figure}

\begin{figure}
\includegraphics[width=14cm]{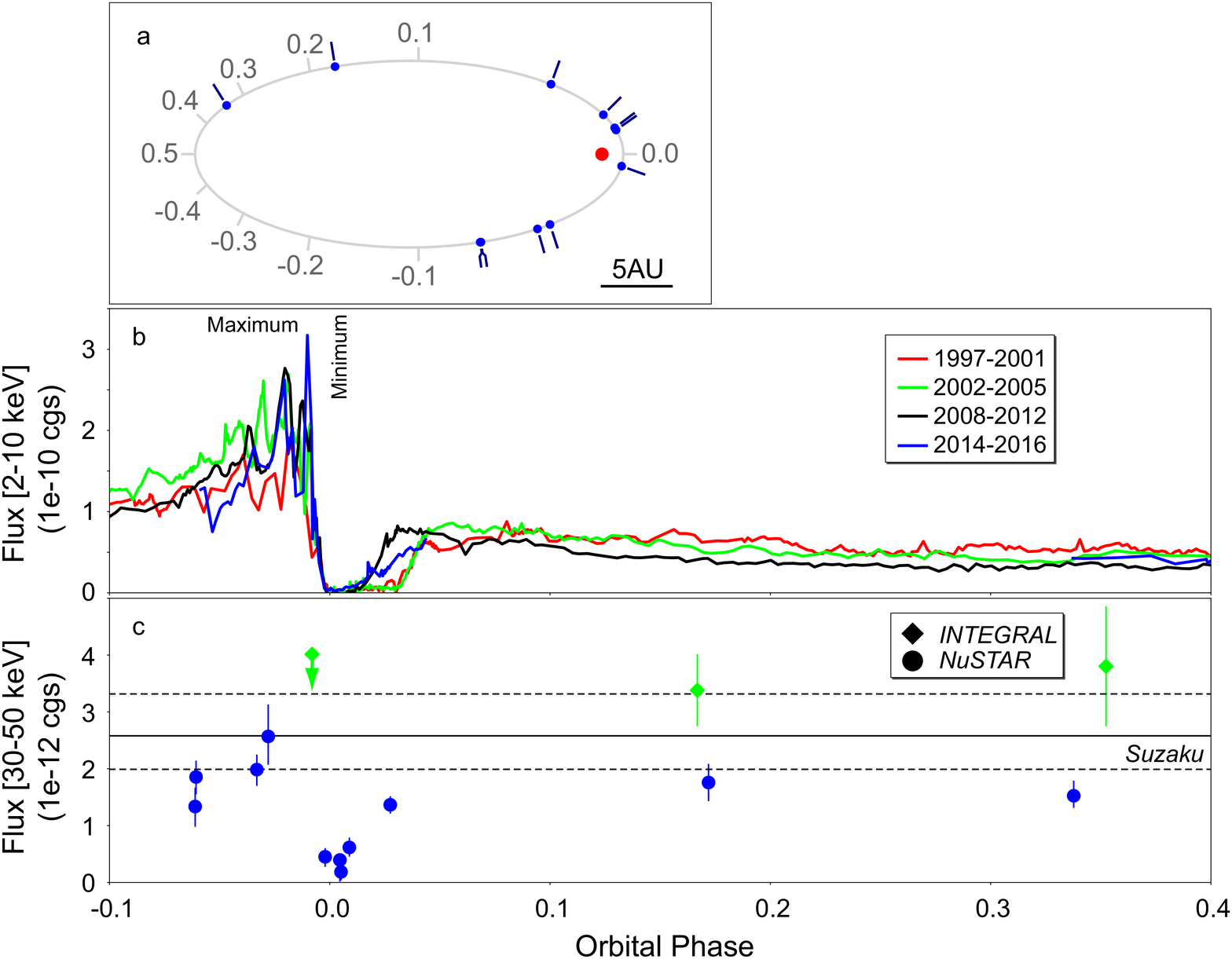}
\caption{
\textbf{Flux variations of the thermal and non-thermal X-ray components with orbital phase.}
{\bf a}, Binary orbital positions of the companion during the \NUS\ observations.
The periastron timing is not constrained better than $\approx$0.02 in phase, so that the actual positions
especially near periastron have large uncertainties.
The companion size is not to scale.
{\bf b}, \RXTE\ and \SWIFT\ light curves of \etacar\ between 2$-$10 keV since 1998 \citep{Corcoran2017a}.
The labels ``Minimum"/``Maximum" show the timings of the soft X-ray minimum/maximum discussed in the text.
{\bf c}, 30$-$50~keV X-ray flux of the flat, power-law component measured with \NUS\ between 2014$-$2016 ({\it blue}), 
assuming a power-law photon index at 1.65.
The solid and dotted black horizontal lines are the best-fit flux and its 90\% confidence range of a power-law component
measured with \SUZAKU\ assuming the flux is constant throughout the orbit \citep{Hamaguchi2014b}.
The \INTEGRAL\ \citep[{\it green diamond,}][]{Leyder2008,Leyder2010} and \SUZAKU\ measurements were 
converted to 30$-$50 keV fluxes.
Error bars are shown at 2$\sigma$.
\label{fig:flux}
}
\end{figure}

The soft X-ray ($<$15~keV) spectra obtained by {\NUS} are characterized by thermal emission from plasma with a maximum 
temperature of 4$-$5~keV (Figure~\ref{fig:spectra}), 
which is consistent with the \XMM\ spectra simultaneously obtained,
and previous analyses of {\etacar}'s thermal X-ray emission \citep[e.g.,][]{Hamaguchi2014a}.
However, the extremely hard ($\gtrsim$15~keV) X-ray emission seen in 2015 and 2016,
following {\etacar}'s periastron passage in 2014, 
is significantly brighter and flatter in slope than the \KT~$\sim$4$-$5~keV plasma emission in this energy range,
and is detected above background up to energies of 50~keV.
The spectrum obtained in 2014 March 31, which is 4 times brighter than the 2015 and 2016 spectra below 15~keV,
follows the \KT $\sim$4.5~keV thermal emission spectrum up to 30~keV,
but it flattens above that energy and converges to the 2015 \& 2016 spectrum.
The other two observations obtained near the maximum of the thermal X-ray emission, 
which occurs just prior to periastron passage (Figure~\ref{fig:flux}),
follow a similar trend in the hard band slope and converge to the 2015 \& 2016 spectrum in the same way.
This result confirms the \KT~$\sim$4$-$5~keV thermal component variability with orbital phase seen previously, 
but it reveals that the highest energy emission is characterized by a flat emission component that is nearly 
constant outside periastron passage.

The \NUS\ spectrum, however, shows that 
this hard flat component nearly disappears during the minimum of the \KT~$\sim$4$-$5~keV thermal emission 
near periastron passage.
This \KT~$\sim$4$-$5~keV thermal X-ray minimum is believed to be
caused by orbital changes in the head-on wind collision
both geometrically (i.e., eclipse by the primary wind) and mechanically (decay of the collisional shock activity) \citep{Hamaguchi2014a}.
The decline of the hard, flat component along with \KT~$\sim$4$-$5~keV thermal X-ray minimum,
as well as the positional coincidence of the extremely hard source with \etacar,
is conclusive proof that \etacar\ itself, and its colliding wind activity, 
is the source of this flat high-energy X-ray component.

If the 30$-$50~keV emission is thermal in nature, it would require a temperature of \KT~$\gtrsim$20 keV, 
a temperature much higher than could be mechanically produced by the wind of either star.
Thus the hard flat source must be produced by non-thermal processes.
We characterize the spectrum using a simple power-law spectrum of the form $KE^{-\Gamma}$ 
(where $K$ is the flux normalization, $E$ the photon energy, and $\Gamma$ the photon index).
We minimized the systematic uncertainty of the instrumental and cosmic background 
through a detailed background study.
Our analysis constrained $\Gamma$ to be less than 3.
Values of $\Gamma \sim$3 can be ruled out since the non-thermal emission would then contribute significantly 
to the observed emission below $10$~keV 
at phases away from periastron; this  would cause a variation of the equivalent width of the strong thermal line from He-like iron at 6.7~keV with phase, which is not seen.
Therefore, the photon index has to be in the range $\Gamma \lesssim2$.

There are several non-thermal emission processes that the colliding wind activity can drive --- synchrotron emission, 
synchrotron self-Compton, IC up-scattering of stellar photons, relativistic bremsstrahlung and pion-decay.
However, to match the observed flux at 50 keV, the synchrotron process would require electrons with 
Lorentz factor $\gamma \sim$3$\times$10$^{6}$ for a reasonable magnetic field strength ($B \sim$ 1~Gauss),
which do not seem likely to exist given the expected strong IC cooling \citep[e.g.,][]{Pittard2006b}.
Pion-decay emission peaks at 67.5 MeV and is important only above $\sim$10 MeV, while relativistic bremsstrahlung emission 
and synchrotron self-Compton are unlikely to match the emission from IC up-scattering \citep[e.g.,][]{Pittard2006}.
Furthermore, the value of $\Gamma \lesssim$2 we derived is typical of 1st order Fermi acceleration and 
similar to the radio indices measured from another well-known massive colliding wind binary system, WR 140 \citep{Dougherty2005}.
Thus IC up-scattering is the most plausible mechanism to produce the non-thermal emission in the extremely hard X-ray band.

This result demonstrates the presence of a high-energy non-thermal X-ray source physically associated with \etacar\
and lends additional strong support to the idea that the $\gamma$-ray source is also physically associated with \etacar.
With the now established physical association between the \NUS\ and \FERMI\ sources, it now makes sense to consider 
a consistent model for both the X-ray and $\gamma$-ray emission.
The extremely hard X-ray component seen by \NUS\ smoothly connects to the soft GeV $\gamma$-ray spectrum at a power-law slope of $\Gamma \sim1.65$
(Figure~\ref{fig:spectra} {\it right}).
This component also shows similar flux variation to the soft GeV component \citep[Figure~\ref{fig:flux} {\it bottom},][]{Reitberger2015}.
These characteristics strongly suggest that
the non-thermal X-ray component seen by \NUS\ is the low-energy tail of the soft GeV $\gamma$-ray component
produced by the IC mechanism \citep{Abdo2010,Farnier2011}.
There would be no obvious connection between the $\gamma$-ray and hard X-ray emission 
if the soft GeV $\gamma$-ray component originates from the pion decay process \citep{Ohm2015}.

Earlier \INTEGRAL\ and \SUZAKU\ flux measurements of extremely high energy emission were 2$-$3 times larger 
than our \NUS\ measurements \citep[Figure~\ref{fig:flux},][]{Leyder2010,Hamaguchi2014b},
but the soft GeV emission has not varied remarkably since the beginning of {\FERMI}'s monitoring in 2008.
This discrepancy either indicates some
cycle-to-cycle variation in the non-thermal emission (which seems unlikely given the consistency of the \NUS\ and \FERMI\ 
spectra), or that these earlier measurements have overestimated the intrinsic source flux 
due to poorly determined backgrounds or other issues.

A puzzle is the lack of an increase in luminosity of this IC scattered component as the thermal plasma emission increases near periastron.
If the non-thermal electrons fill the wind colliding region,
the IC luminosity should be proportional to the product of 
the number of non-thermal electrons and the intensity of the stellar UV,
and the product is also proportional to the thermal plasma luminosity for a constant temperature.
That this variation is not observed can be explained by the rapid cooling that the non-thermal electrons undergo due to IC scattering as the stars approach each other.
Because of this effect, the non-thermal electrons that are capable of producing 50 keV photons ( i.e. those with a Lorentz factor $\gamma \sim$200)
gradually exist only in a thin layer downstream of the shock \citep{Pittard2006b}, rather than filling the entire wind colliding region.
This process would decrease the number of non-thermal electrons near periastron
and produce a flat light curve toward the X-ray maximum.

By localizing the position of the high energy source to better than 5\ARCSEC, and by showing that the source varies in phase 
with the lower-energy X-ray emission, our \NUS\ observations prove conclusively that \etacar\ is clearly a source of 
non-thermal high-energy X-ray emission,
and connect the non-thermal X-rays to the soft GeV $\gamma$-ray source detected by \FERMI.
This confirms that a colliding wind shock can accelerate particles to sub-TeV energies.
Since the colliding-wind shock occurs steadily, persistently, and predictably,
massive binary systems are potentially important systems for studying particle acceleration by the Fermi process in an astrophysical setting.
The emission we observe is consistent with IC upscattering of lower-energy stellar photons. 
IC emission should also be accompanied by lower-energy synchrotron emission, which has not been detected.
However, synchrotron emission from \etacar\ would be difficult to detect because of strong thermal dust emission from the surrounding nebula,
and because a suitable high-spatial-resolution radio interferometer in the southern hemisphere is not yet available.
The Square Kilometer Array, which is under construction in South Africa, 
may eventually detect this emission component from \etacar.
Although there are other massive binary systems with strong colliding wind shocks, such as WR~140,
only \etacar\ has been confirmed as a $\gamma$-ray source.
Studying the differences amongst these systems in their X-ray and $\gamma$-ray emission will help elucidate the particle acceleration mechanism.

\begin{addendum}
 \item This research has made use of data obtained from the High Energy Astrophysics Science Archive
Research Center (HEASARC), provided by NASA's Goddard Space Flight Center.
This research has made use of NASA's Astrophysics Data System Bibliographic Services.
We appreciate Drs. M. Yukita, K. Madsen and M. Stuhlinger on helping resolve the \NUS\ and \XMM\ data analysis.
K.H. is supported by the \CHANDRA\ grant GO4-15019A, GO7-18012A, the \XMM\ grant NNX15AK62G, NNX16AN87G, NNX17AE67G, NNX17AE68G,
and the ADAP grant NNX15AM96G.
C.M.P.R. was supported by an appointment to the NASA Postdoctoral Program at the Goddard Space Flight Center, 
administered by Universities Space Research Association under contract with NASA.
A.F.J.M. is supported by NSERC (Canada) and FQRNT (Quebec).

\item[Author Contributions] 
K.H. and M.F.C. led the project, from proposing and planning observations, analyzing the data to composing the manuscript.
J.M.P. constructed a theoretical model that explains the variation of the non-thermal component.
N.S. performed initial analysis of the \NUS\ data in 2015.
H.T analyzed and discussed \FERMI\ data of \etacar.
C.M.P.R. performed theoretical simulations of {\etacar}'s thermal X-ray emission.
B.W.G. and D.R.W. discussed \NUS\ data analysis, especially the background characteristics.
T.R.G. worked for the observation planning.
T.R.G., N.D.R., T.I.M., and A.F.J.M. discussed the wind property of \etacar.
All authors reviewed the manuscript and discussed the work.

 \item[Competing Interests] The authors declare that they have no
competing financial interests.
 \item[Correspondence] Correspondence and requests for materials
should be addressed to K.H.~\\(email: kenji.hamaguchi@nasa.gov).
\end{addendum}


\clearpage

\begin{center}
{\LARGE Methods}
\end{center}

\noindent{\large\bf \NUS\ Data}\\
\noindent{\bf Observations:}
\NUS\ has two nested Wolter I-type X-ray telescopes with a 2$\times$2 array of CdZnTe pixel detectors
in each focal plane module \citep[FPMA/FPMB, ][]{Harrison2013}.
These mirrors are coated with depth-graded multilayer structures and focus X-rays over a 3$-$79 keV bandpass.
They achieve an angular resolution of roughly 60\ARCSEC\ half power diameter \citep{Madsen2015a}.
The focal plane detectors are sensitive between 3$-$79~keV and cover a 12\ARCMIN\ \FOV.
The energy resolution of the detectors is 400 eV below $\sim$40 keV, rising to $\sim$ 1 keV at 60 keV.
Stray light contamination is not an issue unless there are bright sources ($>$100~mCrab) within 1\DEGREE\ to 5\DEGREE\ of the target.

\NUS\ observed \etacar\ on 9 occasions and produced 11 datasets with different observation identifiers (ObsID).
Two datasets on 
2014 March 31 (ObsIDs: 30002010002, 30002010003) and 
2014 August 11 (ObsIDs: 30002010007, 30002010008) 
were performed consecutively, but they have different ObsIDs due to small pointing offsets.
The list of the datasets is summarized in Supplementary Table~1.
We used the HEASoft package\footnote{https://heasarc.nasa.gov/lheasoft/}, version 6.20 or above,
to analyze the \NUS\ data.

\noindent{\bf Reduction and Accurate Measurement of the \NUS\ Background:}
Measuring the spectrum of \etacar\ at energies above 10 keV requires some care.  At the lower end of this energy range, emission is significantly affected by the high-energy tail of {\etacar}'s thermal source at a temperature of $\sim4.5$~keV, 
and which we were able to precisely measure using \XMM\ X-ray spectra in the 2$-$10~keV energy range.  
At higher X-ray energies, the thermal contribution is negligible 
(except for a short interval during the 2$-$10~keV X-ray maximum just before periastron),
but instrumental and cosmic background components grow in importance.
Our analysis requires careful measurements of {\etacar}'s spectral shape above $\sim$25~keV, 
where non-thermal emission exceeds  \KT~$\sim$4.5~keV thermal emission.
X-ray emission from \etacar\ in this energy band is weak and comparable to \NUS\ particle background.
Therefore, we maximized the source signal with respect to background by
i) removing high background intervals during each observation, and
ii) employing a small source region.
We then accurately estimated the background spectrum by utilizing the background estimate tool {\tt nuskybgd} \citep{Wik2014a}.

Background particle events of the \NUS\ detectors sometimes increase abruptly when \NUS\ is near the South Atlantic Anomaly (SAA).
After reviewing the background variation in each observation\footnote{http://www.srl.caltech.edu/NuSTAR\_Public/NuSTAROperationSite/SAA\_Filtering/SAA\_Filter.php},
we removed the high background intervals with the tool {\tt saacalc} using the option, {\tt saacalc=2 --saamode=optimized --tentacle=yes}.
In all observations with abrupt background increases, this option removed high background intervals, 
by decreasing  exposure times by $\lesssim$5\%.
This process significantly reduced background of NUS$_{\rm 160615}$ by $\approx$40\% between 30$-$60 keV.

For extracting source light curves and spectra from each dataset, we used a circular region with a 30\ARCSEC\ radius,
which includes $\sim$50\% of the X-ray photons of an on-axis point source.
Since the source region is comparable to the mirror point-spread-function (\PSF) size and there is a positional offset 
in the absolute coordinates and the coordinate systems between FPMA and FPMB by up to $\sim$10\ARCSEC,
we re-calibrated the absolute coordinates on each detector image frame from a two-dimensional image fit with a \PSF\ image.
\CHANDRA\ observations indicate that colliding wind emission from \etacar\ dominates the emission below 10~keV,
so that we measured the peak position of \etacar\ between 6$-$8~keV in each detector image
by an on-axis \PSF\ with the \CHANDRA\ software {\tt CIAO/Sherpa}.
Before each fit, the \PSF\ image was rotated to consider the satellite roll angle.

We then measured the \NUS\ background from surrounding source-free regions using {\tt nuskybgd}.
This tool extracts spectra from specified source-free regions and fits them simultaneously for
known background components --- line and continuum particle background, cosmic X-ray background (CXB)
passing through the mirror (focused) and unblocked stray light in the detector (aperture), and solar X-rays
reflecting at the mast. For the \etacar\ data, we ignored the solar reflection component as it is very soft ($\lesssim$5~keV).

There are a few more components that we added in the {\tt nuskybgd} model for the \etacar\ data (see Supplementary Figure~2).
One is the Galactic Ridge X-ray Emission (GRXE).
As \etacar\ is located almost on the Galactic plane ($l, b$) = (287.6\DEGREE, $-$0.63\DEGREE),
GRXE from \KT\ $\sim$6~keV thermal plasma is as strong as CXB at $\sim$7 keV \citep[e.g.,][]{Miyaji1998}.
This emission comes from both the mirror and opening between the mirror and focal plane modules (stray light) similar to the CXB.
The only difference is that GRXE is concentrated within $\sim$4\DEGREE (FWHM) from the Galactic plane \citep[e.g.,][]{Valinia1998},
while CXB is uniform on the sky.
Earlier measurements give good estimate of the two (focused \& aperture) CXB components and focused GRXE.
We thus measured the contribution of aperture GRXE contamination by fixing the parameters for the other sky background components.
For this measurement, we used 3 datasets obtained during the lowest soft X-ray flux phase 
(NUS$_{\rm 140728}$, NUS$_{\rm 140811a}$, NUS$_{\rm 140811b}$)
since \etacar\ outshines the entire detector \FOV\ outside the soft X-ray minimum.
X-ray emission from unresolved young stars in the Carina nebula is not negligible below $\sim$7~keV,
so that we fit the background spectra only above this energy range.
We assume the GRXE spectral shape is similar to that in \citep{Ebisawa2005}, which is measured for GRXE 
at ($l, b$) = (28.5\DEGREE, 0.0\DEGREE), but we changed its normalization to match the GRXE flux at the \etacar\ position \citep{Miyaji1998}.
We extracted data from 4 source regions, each of which has 5.5\ARCMIN$\times$5.5\ARCMIN, each of which covers a detector (0, 1, 2, 3) on each module
(FPMA/FPMB), excluding areas around the bright hard X-ray sources, \etacar, WR~25, and HD~93250.
This analysis shows that the observed stray light flux is 82\% (FPMA) and 75\% (FPMB) of the expected stray light
if the GRXE has the same surface brightness as at ($l, b$) = (285\DEGREE, 0.0\DEGREE).
We fixed the GRXE contamination at these values for the rest of the background analysis.
These ratios may change with the satellite roll angle, but our conclusions  should not be significantly affected 
as the GRXE is negligible above 15~keV.

The other background component accounts for particle background variations between the detectors.
{\tt Nuskybgd} assumes that instrumental background is uniform between the detectors (0, 1, 2, 3), but 
some \NUS $>$15~keV images of \etacar\ show small but significant fluctuations (see Supplementary Figure~1).
These fluctuations possibly originate from the sensitivity difference between the detectors (private comm. Kristin Madsen),
or Cen X-3 contamination through the detector light baffle.
In either case, these fluctuations can introduce up to $\sim$10\% normalization error at the \etacar\ position in some observations.
We therefore added a contamination component to the {\tt nuskybgd} model, an absorbed power-law model 
({\tt TBabs $\times$ Power-law}) whose normalization was allowed to vary between the detectors;
the normalization for the detector with the lowest enhancement was fixed at zero.
We added this component to the background model for \etacar.

Using these constraints, we ran {\tt nuskybgd} to estimate background for all \etacar\ datasets.
Since we need a precise measurement of the background above 25~keV,
we used a larger region for each detector to increase the photon statistics
--- the region includes WR~25 and HD~93250, which have little flux above 15 keV ---
and excludes smaller areas around \etacar.
We fit the unbinned estimated background spectra above 15 keV up to 150~keV using Poisson statistics to give the best measurement 
of the estimated background shape between 25$-$79~keV.
We then normalized the best-fit result for each \etacar\ spectrum.

The background subtracted spectrum and the corresponding simulated background spectrum for 
each observation is shown in the Supplementary Figure~3.
Three spectra shown in Figure~2a are co-additions of the spectra 
NUS$_{\rm 140331a}$ and NUS$_{\rm 140331b}$ ({\it black}),
NUS$_{\rm 150716}$ and NUS$_{\rm 160615}$ ({\it red}), and 
NUS$_{\rm 140811a}$ and NUS$_{\rm 140811b}$ ({\it orange}).
For spectral fits, we add the normalized background model to the source model and 
fit the source spectra using Poisson statistics.

\noindent{\bf Analysis: }
As described in the previous section, the absolute coordinates on each image have uncertainties of several arc-seconds.
For Figure~1, we shifted each detector image by pixel offsets measured with the \PSF\ fits to 6$-$8~keV images 
and combined them for each band.
We recalibrated the absolute coordinates based on the soft band image.
We smoothed the image with a Gaussian of $\sigma =$8~pixels to increase the photon statistics.
Supplementary Figure~1 also shows the entire field of view of the co-added \NUS\ images of
NUS$_{\rm 150716}$ and NUS$_{\rm 160615}$.

The X-ray spectrum of \etacar\ is complex with these components which contribute to the emission above 3 keV:
i) variable multi-temperature thermal components produced by the hot, shocked colliding wind plasmas;
ii) a weak, stable central constant emission (CCE) component, which probably originates from hot shocked gas
inside the cavity of the secondary star's wind, which was ejected in the last few orbital cycles;
iii) X-ray reflection from the bipolar Homunculus nebula;
iv) a power-law component with photon index $\Gamma \lesssim2$.
We included all these components in the spectral model, to determine the non-thermal flux variation with orbital phase.

Component i) varies slowly with the binary orbital motion.
Earlier spectral analyses of \etacar\ between 0.5$-$10~keV \citep[e.g.,][]{Hamaguchi2007b}
show that this component can be described with two-temperature components having \KT\ $\sim$4.5 and $\sim$1.1~keV,
each of which suffers independent absorption.
The \NUS\ spectra cannot constrain parameters of the cool (\KT $\sim$1.1 keV) component well
without sensitivity below 3~keV where the emission dominates.
We therefore fixed \KT, elemental abundance and \NH\ of the cool component
at 1.1~keV, 0.8 solar, 5$\times$10$^{22}$~\UNITNH, 
the best-fit values of the \XMM\ EPIC spectra on 2015 July 16.
On the other hand, we allowed parameters of the hot component 
(\KT, abundance, normalization and absorption) to vary in all spectral fits.

Component ii) probably originates from the collision of secondary stellar winds with the primary winds
ejected in early cycles \citep[e.g.,][]{Hamaguchi2007b,Madura2013,Russell2016a}.
This component can be seen in \etacar\ spectra only around the soft X-ray minimum and
it does not change significantly in the latest 3 minima (2003, 2009 and 2014).
This component cannot be observed during other orbital phases, but a theoretical simulation suggests 
that it is stable outside of the minimum as well \citep{Russell2016a}.

Component iii) originates from the reflection of the colliding wind X-ray emission at the surrounding Homunculus bipolar nebula.
The variation follows the wind colliding emission from the central binary system, with light travel time-delay
by 88 days, on average \citep{Corcoran2004}.
This component is extended ($\sim$20\ARCSEC) and can be spatially resolved with \CHANDRA.
This component is weaker than the CCE (Component ii) except for the Fe fluorescence at 6.4~keV.
We therefore fixed this component to the best-fit spectrum derived from the \SUZAKU\ observation
during the deep X-ray minimum phase in 2014 \citep{Hamaguchi2016a}.
The components (ii) + (iii) only contribute $\sim$10\% to the spectra after the recovery in 2015 and 2016,
and dominate during the X-ray minimum.

Component iv) is proved to be present from the \NUS\ observations in this paper.
It dominates emission above 30 keV, and does not vary significantly outside the soft X-ray minimum.
No spectra show the shape of this component below 30~keV clearly.
However, our measurement of the equivalent width of the He-like iron K line varies less than 10\%
through the orbit outside of the X-ray minimum.
This means that the non-thermal component is less than 10\% of the thermal continuum at 6.7 keV,
which constrains the photon index at $\Gamma<$2.
We choose $\Gamma =$1.65 for consistency between the \NUS\ and \FERMI\ data, 
but the conclusions we draw do not change significantly for $\Gamma \lesssim$2.
The absorption column for the power-law component is tied to that of the hot \KT\ component.
This is based on the assumption that the non-thermal emission originates from the apex of the colliding wind region,
but changing this \NH\ does not affect the fitting result for $\Gamma<$2.

We simultaneously fit unbinned \etacar\ spectra of both focal plane modules (FPMA, FPMB) 
using the maximum likelihood method assuming Poisson statistics (c-stat in Xspec).
The normalizations of the spectral models between FPMA and FPMB are independently varied to consider small effective area calibration uncertainty.
The errors are estimated using  Markov Chain Monte Carlo simulations (mcmc in Xspec).
The fitting results are shown in Figure~3 and Supplementary Table~2.

\noindent{\large\bf \XMM\ Data}

\noindent{\bf Observations:}
\XMM\ has three nested Wolter I-type X-ray telescopes \citep{Aschenbach2000} with 
the European Photon Imaging Camera (EPIC) CCD detectors
(pn, MOS1 and MOS2) in their focal planes \citep[][]{Struder2001, Turner2001}.
They achieve a spatial resolution of 15\ARCSEC\ half power diameter
and an energy resolution of 150~eV at 6.4~keV\footnote{http://xmm-tools.cosmos.esa.int/external/xmm\_user\_support/documentation/uhb/XMM\_UHB.pdf}.
There are three \XMM\ observations simultaneous with the \NUS\ observations, 
two of which are reported in \citep{Hamaguchi2016a}.
In all observations,
the EPIC-pn and MOS1 observations were obtained in the small window mode with the thick filter to avoid photon pile-up and optical leakage,
though the EPIC-MOS1 data in XMM$_{\rm 140606}$ was still affected by photon pile-up.
The EPIC-MOS2 observations used the full window mode with the medium filter to monitor serendipitous sources around \etacar,
so that its \etacar\ data are significantly affected by photon pile-up and optical leakage and 
thus provide no useful information about \etacar.
Fortunately, most of the \XMM\ observations were obtained during periods of low particle background.

\noindent{\bf Analysis:}
We followed \citep{Hamaguchi2007b} for extracting  \XMM\ source spectra,
taking the \etacar\ source region from a 50\ARCSEC$\times$37.5\ARCSEC\ ellipse 
with the major axis rotated from the west to the north at 30\DEGREE.
For background, we used regions with negligible emission from \etacar\ on the same CCD chip.
In addition, we limited the EPIC-pn background regions using nearly the same RAWY position of \etacar,
according to the \XMM\ analysis guide\footnote{http://xmm.esac.esa.int/sas/current/documentation/threads/PN\_spectrum\_thread.shtml}.
The source did not show significant variation.
We assumed chi-square statistics for the \XMM\ fits to the background-subtracted spectra.

The \XMM\ spectra show multiple emission lines, notably from helium-like Fe K emission lines.
The Fe K emission line is shifted by $\sim$25~eV for both EPIC-pn and MOS1,
which corresponds to $v \sim$1100 \UNITVEL.
However, the simultaneous \NUS\ observation did not show such a shift,
and a \CHANDRA\ HETG grating observation of \etacar\ obtained at a very similar orbital phase, but one cycle previously
(ObsID: 11017, 11992, 12064, 12065, Date: 2009 Dec 21$-$23, $\phi_{\rm obs}$ = 2.168)
gives only a small shift of $\sim$7~eV.
In addition, we saw a similar energy shift in \XMM\ data obtained with the same observing mode in 2014.
The shift seen in the \XMM\ spectra is probably due to an error in energy-scale calibration.

After adjusting the gain shift,
the \XMM\ spectra of \etacar\ are successfully reproduced by a model with the cooler \KT\ at 1.1 keV and hotter \KT\ at 4.5 keV.
These temperatures are similar to those measured in early \XMM\ observations \citep{Hamaguchi2007b}.

\noindent{\large\bf Theoretical Model for the Constancy of the Non-thermal Component}

If the non-thermal electrons fill the wind-colliding region, 
the IC luminosity, $L_{\rm IC}$, should be proportional to the number of non-thermal electrons
($N_{\rm acc} \propto nV$, where $n$ and $V$ are respectively the number density of the thermal plasma in the wind colliding region 
and the volume of the wind colliding region) and the intensity of the stellar UV ($U_{\rm UV}$).
Since $n$ and $U_{\rm UV}$ are both $\propto D^{-2}$, and $V \propto D^{3}$, we might expect $L_{\rm IC} \propto 1/D$, 
where $D$ is the stellar separation.
Therefore, the $L_{\rm IC}$ should follow the same variation as the X-ray luminosity of the thermal plasma (i.e. 2$-$10~keV light curve in Figure~3b), 
which also has the $1/D$ dependence valid for the adiabatic limit \citep{Stevens1992}.

That this variation is not observed can be explained by the rapid cooling that the non-thermal electrons undergo due to IC scattering as they flow downstream from the companion star's shock\footnote{
For particles to be accelerated the shocks must be collisionless and mediated by the magnetic field. This requires that the postshock thermal collision timescale must be longer than the ion gyroperiod. This is not satisfied at high densities \citep[see, e.g.,][]{Eichler1993}. Since the shocked luminous blue variable wind is highly radiative, its post-shock density is several orders of magnitude greater than the post-shock density of the companion's wind, and is not likely to be collisionless.}.
Rather than filling the entire wind colliding region, the non-thermal electrons which are capable of producing 50 keV photons 
(those with a Lorentz $\gamma \sim$200) instead only exist in a thin layer downstream from the shock \citep{Pittard2006b}.
For reasonable values (e.g. $D$ = 10 au, $r_{\rm O}/D$ = 0.3, where $r_{\rm O}$ is the distance from the companion star to 
the shock on the line-of-centres, 
$L_{\rm UV}$ = 5$\times$10$^{6}$ \UNITSOLARLUMI) the rate at which the non-thermal electrons lose energy due to IC scattering is $d{\gamma}/dt$ $\sim$10$^{-6}$~$\gamma^{2}$ s$^{-1}$ 
\citep[cf. Eq. 4 in][]{Pittard2006b}.
Hence it takes roughly 6000 second (= $t_{\rm cool}$) to cool from the expected maximum energy of the electrons at the shock ($\gamma \sim$ 10$^{5}$) 
to $\gamma \sim$200. 
During this time the electrons will have travelled downstream from the shock a distance of $d_{\rm cool} = v_{\rm ps} t_{\rm cool}$,
where $v_{\rm ps}$ is the post-shock wind velocity.
Using $v_{\rm ps} = v_{\rm wind O}/4$ (appropriate for the gas on the line-of-centres between the stars), the cooling length $d_{\rm cool} \sim 0.01~D$. 
This sets the thickness of the region where non-thermal electrons are capable of producing 50 keV photons.
As the stars approach each other, IC cooling becomes stronger and stronger, and $d_{\rm cool}/D$ decreases. 
Since $d\gamma/dt \propto D^{-2}$, $d_{\rm cool}/D \propto D$. 
So rather than the volume of non-thermal emitting particles scaling as $D^{3}$, it instead scales as $D^{4}$ ($D^{2}$ from the surface area of the shock(s), 
and $D^{2}$ from the cooling length). 
Hence $L_{\rm IC}$ becomes independent of $D$, as is indeed observed outside of the minimum. 
At some very large value of $D$, $d_{cool}$ will be large enough that the non-thermal electrons completely fill the volume of the wind colliding region, 
at which point $L_{\rm IC}$ should scale as 1/$D$, as originally hypothesized. 
However, this is likely to require a value for $D$ which far exceeds the apastron separation in \etacar. 
If $\gamma>$200, non-thermal electrons are confined to only part of the wind-colliding region, 
and a change in the spectral shape of the non-thermal emission with $D$ is not expected.
So this model naturally explains the constant intensity and spectral shape of the IC emission outside of the X-ray minimum.

\noindent{\large\bf Data Availability}

The raw data of the \NUS\ and \XMM\ observations are available from the NASA HEASARC archive https://heasarc.gsfc.nasa.gov.

\renewcommand{\figurename}{Supplementary Figure}
\renewcommand{\tablename}{Supplementary Table}

\begin{table}[h]
\begin{center}
\begin{footnotesize}
\caption{\textbf{Logs of the \NUS\ Observations}}\label{tbl:obslogs}
\begin{tabular}{lccccc}
\hline\hline
Abbreviation&Observation ID&Observation Start&$\phi_{\rm X}$&Exposure&Duration\\
&&&&(ksec)&(ksec)\\ \hline
NUS$_{\rm 140331a}$& 30002010002& 2014 March 31 06:56& 2.9389& 28.8& 50.1\\
NUS$_{\rm 140331b}$& 30002010003& 2014 March 31 21:26& 2.9393& 49.7& 90.6\\
NUS$_{\rm 140526}$& 30002010005& 2014 May 26 11:21& 2.9669& 79.4& 131.8\\
NUS$_{\rm 140606}$& 30002040002& 2014 June 06 10:31& 2.9721& 32.9& 50.6\\
NUS$_{\rm 140728}$& 30002040004& 2014 July 28 10:31& 2.9979& 61.3& 102.1\\
NUS$_{\rm 140811a}$& 30002010007& 2014 August 11 05:36& 3.0046& 31.0& 61.7\\
NUS$_{\rm 140811b}$& 30002010008& 2014 August 11 23:01& 3.0051& 56.9& 111.3\\
NUS$_{\rm 140819}$& 30002010010& 2014 August 19 16:41& 3.0089& 54.5& 97.1\\
NUS$_{\rm 140926}$& 30002010012& 2014 September 26 00:41& 3.0275& 81.2& 143.2\\
NUS$_{\rm 150716}$& 30101005002& 2015 July 16 01:31& 3.1719& 23.6& 38.7\\
NUS$_{\rm 160615}$& 30201030002& 2016 June 15 02:36& 3.3377& 69.3& 120.0\\ \hline
\end{tabular}
\begin{minipage}{13.5cm}
~~\\
Abbreviation: Abbreviation adopted for each observation. 
Observation ID: Observation identification number of each observation.
Observation Start: Time of the observation start.
$\phi_{\rm X}$: Phase at the center of the observation in the X-ray ephemeris in \citep{Corcoran2005}, $\phi_{\rm X}$ = (JD[observation start] $-$ 2450799.792)/2024.
Exposure: Exposure time excluding the detector deadtime.
Duration: Duration of the Observation. 
\end{minipage}
\end{footnotesize}
\end{center}
\end{table}

\clearpage
\begin{table}[h]
\begin{center}
\begin{footnotesize}
\caption{\textbf{Best-fit Parameters of the Spectral Fits}}\label{tbl:specfitlog}
\begin{tabular}{lccccc}
\hline\hline
Abbreviation&Absorption&\multicolumn{3}{c}{Thermal Plasma (Hot Component)}&$\Gamma$=1.65 Power-law\\ \cline{3-5}
& \NH & \KT & Abundance & Flux[10$-$15~keV]& Flux[30$-$50~keV]\\
&(10$^{23}$~\UNITNH)&(keV)&(\UNITSOLARABUND)&(10$^{-11}$~\UNITFLUX)&(10$^{-11}$~\UNITFLUX)\\ \hline
NUS$_{\rm 140331a}$&1.7~(1.3$-$2.0)&4.2~(4.1$-$4.4)&0.50~(0.47$-$0.53)&1.6&0.13~(0.10$-$0.17)\\
NUS$_{\rm 140331b}$&1.4~(1.2$-$1.6)&4.2~(4.1$-$4.3)&0.58~(0.56$-$0.60)&1.9&0.19~(0.15$-$0.21)\\   
NUS$_{\rm 140526}$& 1.9~(1.8$-$2.1)&4.2~(4.1$-$4.2)&0.58~(0.56$-$0.60)&2.8&0.20~(0.17$-$0.22)\\   
NUS$_{\rm 140606}$& 2.4~(2.2$-$2.6)&4.7~(4.6$-$4.8)&0.55~(0.53$-$0.58)&3.3&0.26~(0.21$-$0.31)\\   
NUS$_{\rm 140728}$& 4.1~(2.6$-$5.8)&4.4~(3.6$-$5.8)&0.40~(0.19$-$0.66)&0.034&0.045~(0.027$-$0.060)\\ 
NUS$_{\rm 140811a}$& 9.7~(9.2$-$55)& 4.5~(fix) & 0.50~(fix) &  $<$0.020&0.039~(0.001$-$0.043)\\
NUS$_{\rm 140811b}$& 15~(13$-$19)& 4.5~(fix) & 0.50~(fix) & 0.045 &0.019~(0.004$-$0.033)\\
NUS$_{\rm 140819}$& 8.0~(7.4$-$9.0)&3.3~(3.0$-$3.5)&0.34~(0.27$-$0.41)&0.16&0.062~(0.046$-$0.079)\\  
NUS$_{\rm 140926}$& 2.7~(2.5$-$2.9)&3.3~(3.2$-$3.4)&0.53~(0.50$-$0.56)&0.45&0.14~(0.12$-$0.15)\\  
NUS$_{\rm 150716}$& 1.6~(1.0$-$2.1)&4.0~(3.8$-$4.2)&0.47~(0.41$-$0.53)&0.43&0.18~(0.14$-$0.21)\\  
NUS$_{\rm 160615}$& 0.9~(0.5$-$1.3)&4.4~(4.2$-$4.5)&0.54~(0.50$-$0.57)&0.41&0.15~(0.13$-$0.18)\\ \hline
\end{tabular}
\begin{minipage}{16.5cm}
~~\\
The parentheses quote 90\% confidence ranges.
The plasma temperatures and abundances for NUS$_{\rm 140811a}$ and NUS$_{\rm 140811b}$ are fixed
as they are not well determined with limited photon statistics.
The values in the flux columns are not corrected for absorption.
\end{minipage}
\end{footnotesize}
\end{center}
\end{table}

\clearpage

\setcounter{figure}{0}

\begin{figure}[h]
\caption{
\textbf{Whole \NUS\ images of the \etacar\ field.} \label{fig:image_whole}}
\includegraphics[width=16.5cm]{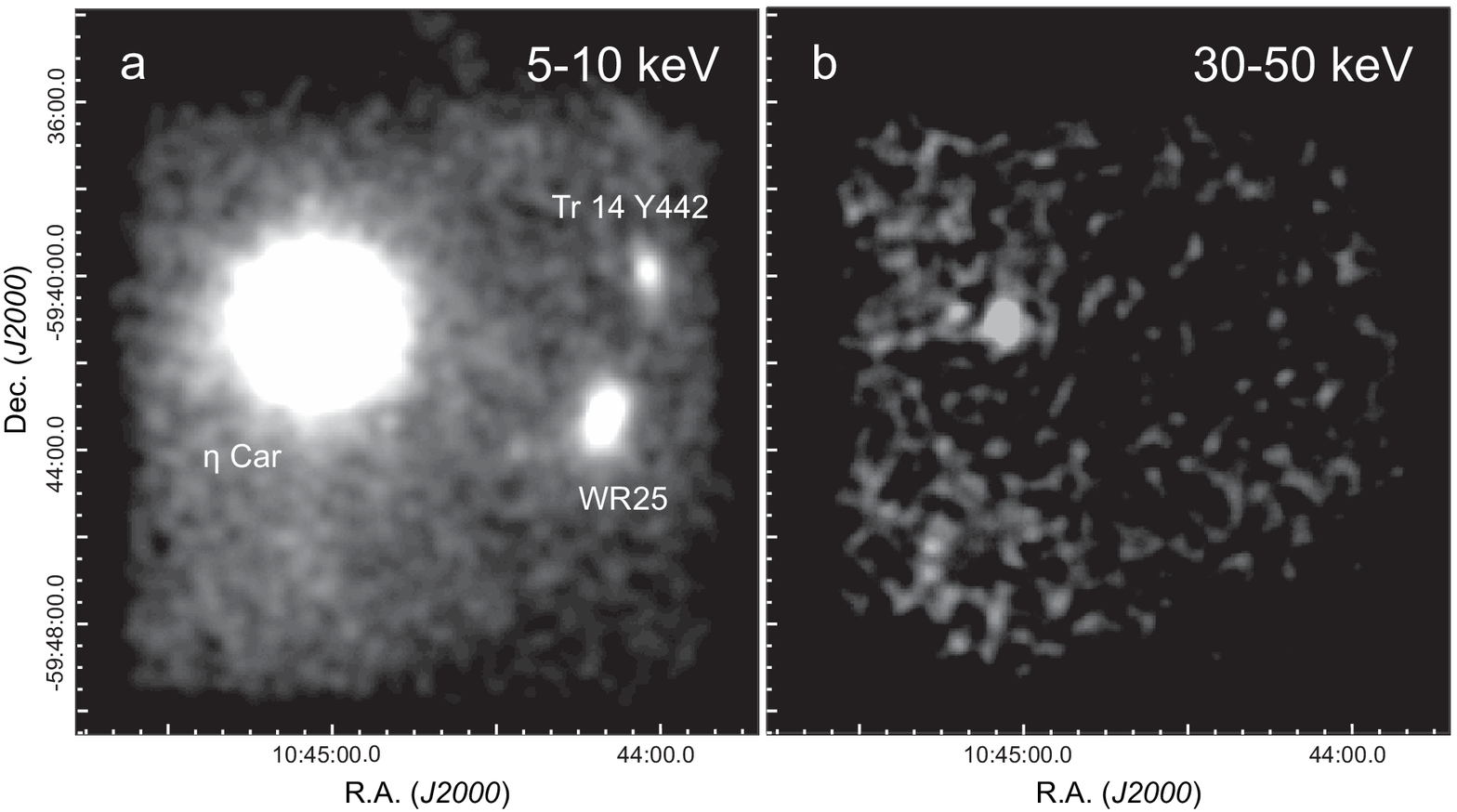}
The images between 5$-$10~keV ({\bf a}, log scale) and 30$-$50 keV ({\bf b}, linear scale)
are produced from the 2015 July 16 ($\phi_{\rm orb} =$0.17) and 2016 June 15  ($\phi_{\rm orb} =$0.34) observations.
Each image is smoothed with a Gaussian function of $\sigma$=8 pixels.
The bright X-ray source in the 5$-$10 keV image, Tr~14 Y442, is a young star in the Carina nebula, which had a giant flare 
during the 2015 observation \citep{Hamaguchi2015a}.
\end{figure}
\clearpage

\begin{figure}[h]
\caption{
\textbf{Individual background components for the FPMA \etacar\ spectrum of the 2016 June 15 observation.}
\label{fig:spec_bgd}
}
\includegraphics[width=16.5cm]{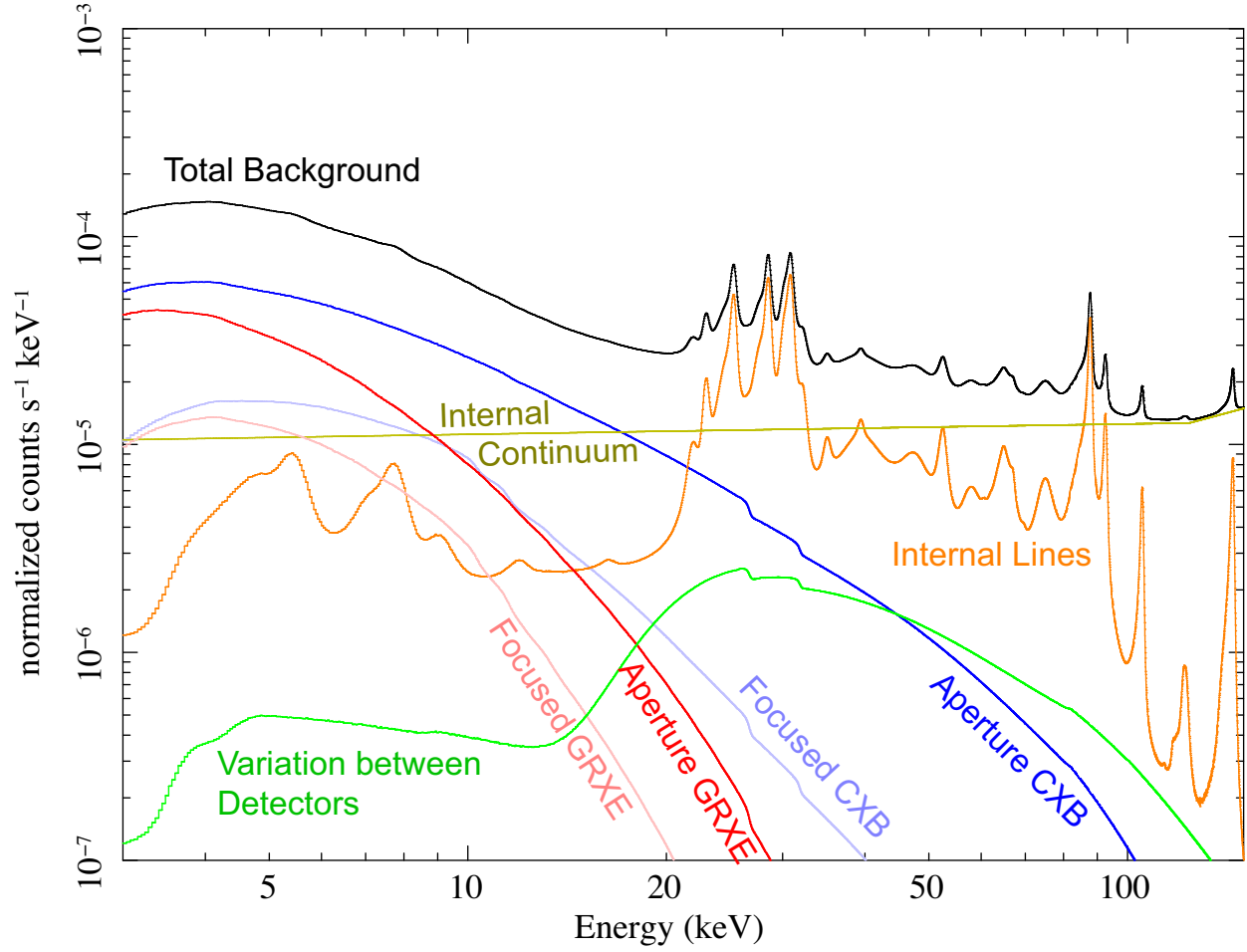}
\end{figure}
\clearpage

\begin{figure}[h]
\caption{
\textbf{Background subtracted \NUS\ FPMA+FPMB spectrum of each observation.}
\label{fig:spec_indiv}
}
\includegraphics[width=15.5cm]{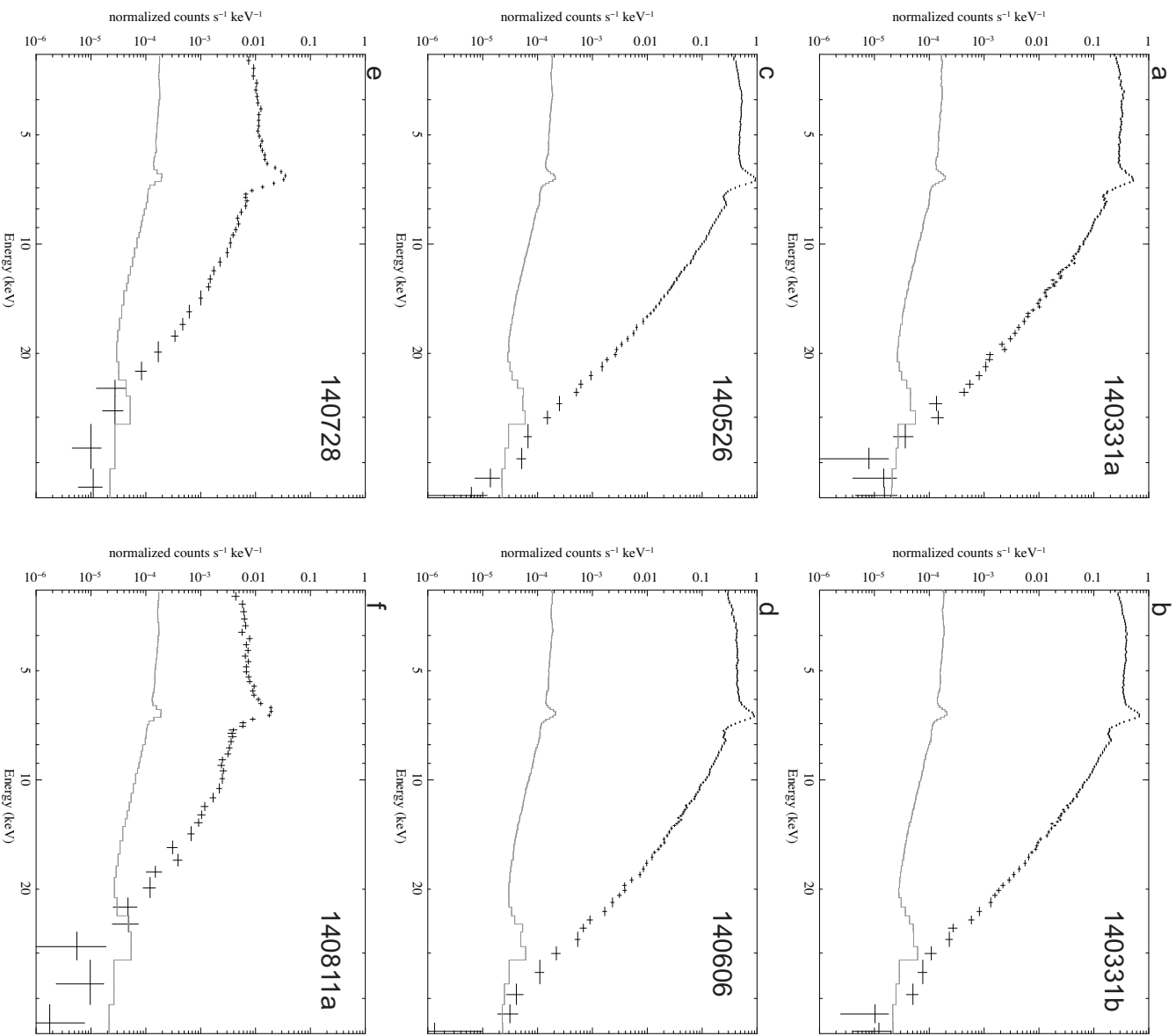}

The solid grey line shows the background level.
\end{figure}

\clearpage

\setcounter{figure}{2}
\begin{figure}[h]
\begin{flushleft}
\caption{Continue.}
\end{flushleft}
\includegraphics[width=15.5cm]{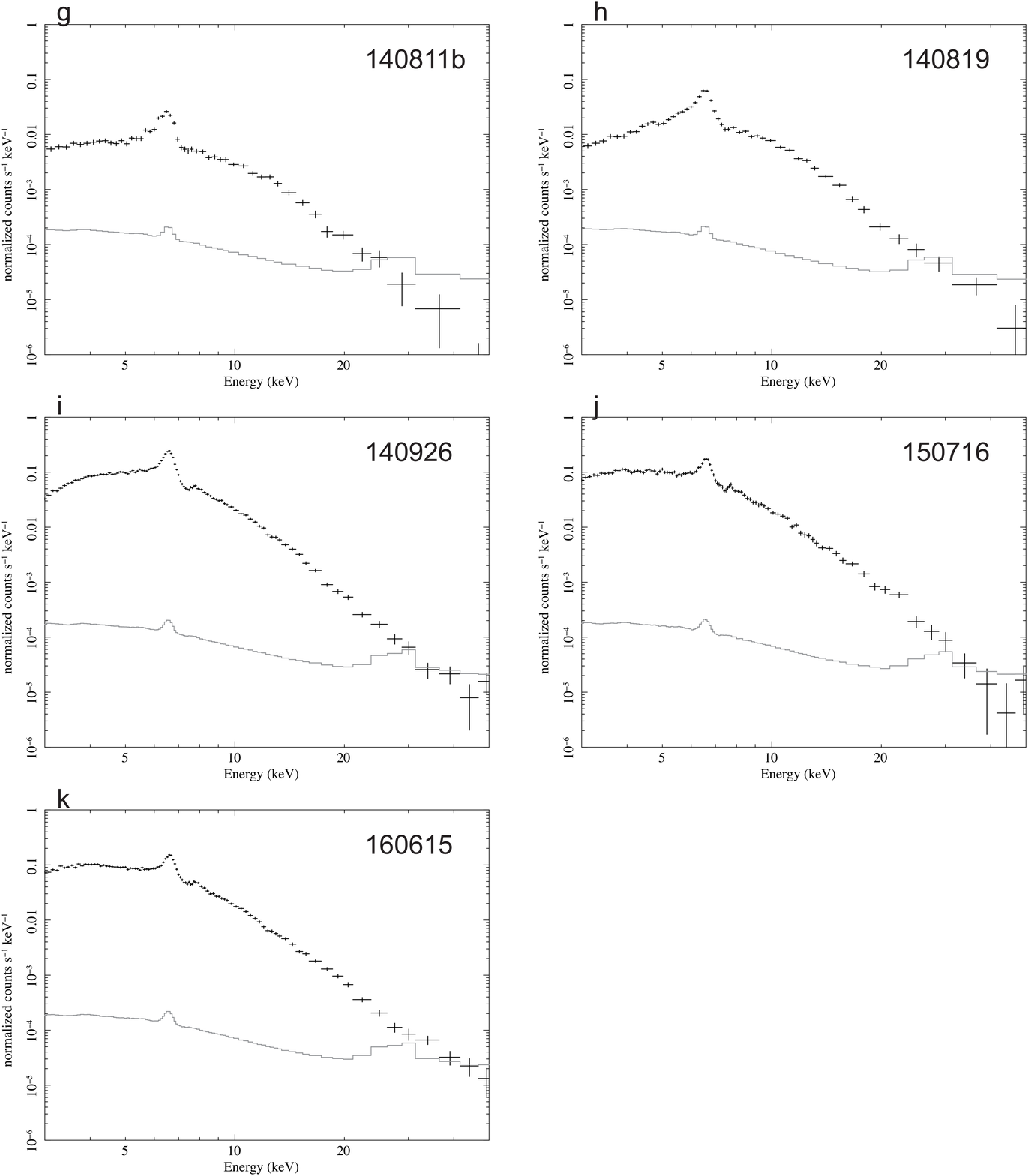}
\end{figure}

\clearpage


\end{document}